\documentclass[onecolumn]{aastex63}

\usepackage[english]{babel}
\usepackage{amsmath,amsfonts,amsthm}
\usepackage{amssymb}
\usepackage{graphicx}

\begin{document}
\title{A Physical Model for the Quasar Luminosity Function evolution between Cosmic Dawn and High Noon}

\author{Keven Ren}
\affiliation{School of Physics, The University of Melbourne, Parkville, Victoria, Australia}
\affiliation{ARC Centre of Excellence for All Sky Astrophysics in 3 Dimensions (ASTRO 3D)}

\author{Michele Trenti}
\affiliation{School of Physics, The University of Melbourne, Parkville, Victoria, Australia}
\affiliation{ARC Centre of Excellence for All Sky Astrophysics in 3 Dimensions (ASTRO 3D)}

\email{kevenr@student.unimelb.edu.au}

% ==== ABSTRACT ==== %
\begin{abstract}

Modeling the evolution of the number density distribution of quasars through the Quasar Luminosity Function (QLF) is critical to improve our understanding of the connection between black holes, galaxies and their halos. Here we present a novel semi-empirical model for the evolution of the QLF that is fully defined after the specification of a free parameter, the internal duty cycle, $\varepsilon_{DC}$ along with minimal other assumptions. All remaining model parameters are fixed upon calibration against the QLF at two redshifts, $z=4$ and $z=5$. Our modeling shows that the evolution at the bright end results from the stochasticity in the median quasar luminosity versus halo mass relation, while the faint end shape is determined by the evolution of the Halo Mass Function (HMF) with redshift. Additionally, our model suggests the overall quasar density is determined by the evolution of the HMF, irrespective of the value of $\varepsilon_{DC}$. The $z\ge4$ QLFs from our model are in excellent agreement with current observations for all $\varepsilon_{DC}$, with model predictions suggesting that observations at $z\gtrsim7.5$ are needed to discriminate between different $\varepsilon_{DC}$. We further extend the model at $z\le4$, successfully describing the QLF between $1\le z\le4$, albeit with additional assumptions on $\Sigma$ and $\varepsilon_{DC}$. We use the existing measurements of quasar duty cycle from clustering to constrain $\varepsilon_{DC}$, finding $\varepsilon_{DC}\sim0.01$ or $\varepsilon_{DC}\gtrsim0.1$ dependent on observational datasets used for reference. Finally, we present forecasts for future wide-area surveys with promising expectations for the Nancy Grace Roman Telescope to discover $N\gtrsim10$, bright, $m_{UV}<26.5$ quasars at $z\sim8$.

\end{abstract}

\section{Introduction}

Decades of quasar observations have led to discoveries of objects out to redshifts $z>7.5$ \citep{Ba_ados_2017, Yang_2020, Wang_2021}, corresponding to a time where the universe was only $\sim 700$Myr old. As highly luminous objects in the early universe, high redshift quasars are crucial tracers for understanding the formation and evolution of early structure. By extension, the quasar luminosity function (QLF), which is a measurement of the luminosity distribution of quasars at a given redshift i.e. a description of the quasar population, can yield information on the joint evolution of black holes, galaxies and their halos. Indeed, the QLF has been utilised in numerous studies to inform and constrain both models of black hole growth \citep{Hirschmann_2014, Sijacki_2015} and galaxy formation \citep{Qin_2017, Marshall_2020}. Observations of the QLF have been enabled by large wide-area surveys across visible and infrared wavelengths, such as the Sloan Digital Sky Survey (SDSS; \citealt{York_2000}), the VISTA Kilo-degree Infrared Galaxy Survey (VIKINGS; \citealt{Peth_2011}), and the Panoramic Survey Telescope and Rapid Response System 1 (Pan-STARRS1; \citealt{Morganson_2012}). These surveys have significantly contributed towards the characterisation of the bright end of the QLF up to $z \sim 6$ \citep{McGreer_2013, Ross_2013, Venemans_2015, Ba_ados_2016, Jiang_2016}. More recently, significant samples of fainter quasars have also been discovered at $z > 4$, by virtue of the wide-area Subaru Strategic Program Survey (Subaru-SSP; \citealt{Akiyama_2017, Matsuoka_2018}) and the Canada-France-Hawaii Telescope Legacy Survey (CFHTLS; \citealt{McGreer_2018}). Remarkably, these recent surveys have allowed, for the first time, measurements of a clear faint end slope beyond the characteristic magnitude, $M^{*}$.

Existing phenomenological models for QLF evolution often assume power-law relations between halo, galaxy and SMBH parameters that are then tuned to match observed correlations at low-$z$ \citep{Wyithe_2003, Conroy_2012}. The advantage of these models is that they provide insight into the more fundamental mechanisms which can result in the observed quasar demographics. However, if we were to simply assess the evolution of the QLF, one can opt for a direct empirical approach instead. An empirical model aims to derive the relationship linking the halo mass function and the quasar luminosity function. The key aspect of these models is that the emergent relation between halo mass and quasar luminosity is independent of our knowledge on the individual mechanisms connecting the properties of halos, its gas and the SMBH. Such empirical models have been extensively utilised in the context of galaxy formation, and have been shown to be robust and reliable in predicting the evolution of properties across cosmic time \citep{Trenti_2010, Behroozi_2013, Mason_2015, Moster_2018}.

In this paper, we take advantage of the recent determinations for the UV QLF to construct a simple semi-empirical model of QLF evolution across $z$. We use the conditional luminosity function (CLF) approach to account for stochasticity, which can be significant in shaping the bright end of the QLF \citep{Ren_2020}, and we also assume that the evolution in the overall quasar density is driven by halo assembly. The free parameters in the model are empirically calibrated using both the $z = 4$ and $z = 5$ UV QLFs from \citet{Akiyama_2017} and \citet{McGreer_2018} respectively. We show that this simple approach well reproduces the QLF at $z = 6$, and is also capable of fitting the QLF at $z \le 4$ with minor amendments. More importantly, this model provides insight into the parameters that guide the evolution of the QLF across cosmic time, thus offering the unique opportunity to inform the physical processes that drive the paradigm between halos, galaxies and the central supermassive black hole (SMBH). Finally, in preparation for future wide-area surveys from the Nancy Grace Roman Space Telescope (RST), we utilise the model to make predictions of the expected number of quasars visible in the observable horizon at higher redshifts.

We first describe the model and its derivation in Section~\ref{sec:model}. We present the modeled QLFs for $z \ge 4$ and $z \le 4$ separately and discuss model implications in Sections~\ref{sec:m4lf} and \ref{sec:l4lf}. In Section~\ref{sec:dc}, we compute the QSO duty cycle as would be inferred from observations by simulating mock surveys of quasars. In Section~\ref{sec:rst}, we use our model to forecast number counts of quasars for future wide-area surveys, such as the proposed wide-field Nancy Grace Roman Telescope (RST) mission. We present MCMC fits for our predicted QLF in Section~\ref{sec:dpl}. Finally, we summarize our work in Section~\ref{sec:conclus}. For this work, we use the WMAP7 \citep{Komatsu_2011} cosmological parameters with $\Omega_{m} = 0.272, \Omega_{b} =0.0455, \Omega_{\Lambda}=0.728, h=0.704, \sigma_{8}=0.81, n_{s}=0.967$. For our analytical halo masses, we use the \citet{Jenkins_2001} halo mass function. The term QLF used in this manuscript implicitly refers to the rest-frame UV QLF (1450\AA) unless specified otherwise.

\section{Model Description\label{sec:model}}

The model builds upon the earlier work of \citet{Ren_2020} suggesting stochasticity in quasar luminosities as the physical link that results in the double power law shape of the observed quasar luminosity function (QLF), in contrast to the Schechter-like form of the underlying halo mass function (HMF). Since stochasticity in the luminosity versus halo mass relation affects the distribution of quasar luminosities hosted by a halo of a given mass, we can also expect stochasticity to play a significant role in the evolution of the QLF based on the changing population of halo masses over time (see \citealt{Ren_2019} for the galaxy LF case).

The construction of our semi-empirical model rests upon two key assumptions: (1) The parameter $\Sigma$, representing the scatter in quasar luminosity as a function of halo mass is an essential component in shaping the bright end of the QLF \citep{Ren_2020}. For simplicity, we assume this to be constant. Even with this simplifying assumption, we will show that the model predictions are consistent with observed QLFs for $z \ge 3$; (2) We include a redshift-dependent parameter that generalises any possible effect of redshift to the quasar luminosity versus halo mass relation. To first order, we assume that this effect is entirely driven by halo growth processes. We choose to characterise this through a parameter defined as the ratio of halo assembly times, $ R_{A}(M_{h}, z_{f}, z_{c}) = [\bar{t}_{z_{c}}(M_{h})/\bar{t}_{z_{f}}(M_{h})]$ between our desired redshift and a calibrated redshift, where $\bar{t}_{z}$ is the median halo assembly time required to assemble a halo of mass $M_{h}$ from $M_{h}/2$ \citep{Giocoli_2007}. As halos typically assemble more rapidly at higher $z$, we expect this ratio to be greater than $1$ for $z_{f} > z_{c}$. To build the QLF, we first define the probability of finding a quasar at a luminosity, $\log L$ within a given halo mass, $M_{h}$ by using a conditional luminosity function (CLF) approach \citep{Ren_2020},

\begin{equation}
\Phi_{z}(\log L \mid M_h)=  (1-\varepsilon_{DC}) \delta(L = 0) + \dfrac{\varepsilon_{DC}}{\sqrt{2\pi}\Sigma}\exp{\bigg( -\dfrac{\log \Big[ {\frac{L}{L_{c}(M_{h}, z, \Sigma, \varepsilon_{DC})}} \Big]^{2}}{2\Sigma^{2}} \bigg) }.
\label{eqn:qclf}
\end{equation}

Here, $L_{c}(M_{h})$ is the median quasar luminosity versus halo mass relation (dropping the variables $z$, $\Sigma$ and $\varepsilon_{DC}$ for brevity), $\Sigma$ is the lognormal scatter in quasar luminosity and $\varepsilon_{DC}$ is the internal quasar duty cycle that is defined as the probability that a halo would contain a black hole that is accreting at any capacity. We clarify that the internal duty cycle does not necessarily correspond directly to the observational duty cycle as determined by measurements of quasar clustering (see Section~\ref{sec:dc}), since quasar clustering can be sensitive to $\Sigma$ \citep{Shankar_2010}. An additional complication is that it is possible for external effects to mimick a change in $\varepsilon_{DC}$. For example, quasar obscuration from the torus or dense neutral gas within the host galaxy \citep{Ni_2020} can be equivalently expressed as a decrease in $\varepsilon_{DC}$. Thus, in practice, $\varepsilon_{DC}$ should be expected to take a value less than unity, $\varepsilon_{DC} < 1$. Finally, we make the assumption that the internal duty cycle is constant in both halo mass and redshift, although we note that a dependence in halo mass can easily be enfolded through Equation~\ref{eqn:qclf}. We will show that these assumptions are adequate for reproducing the QLFs at $z \ge 2$. The CLF is related to the standard luminosity function by,

\begin{equation}
\phi_{z}(\log L) = \int^{\infty}_{0} \dfrac{dn}{dM_{h}}\Bigl |_{z} \Phi_{z}(\log L \mid M_{h}) dM_{h},
\label{eqn:lf}
\end{equation}

where $\frac{dn}{dM_{\rm{h}}}$ is the halo mass function at redshift $z$. In our model, the QLF at a different redshift, $z_{f}$ is given as,

\begin{equation}
\phi_{z_{f}}(\log L) = \int^{\infty}_{0} \dfrac{dn}{dM_{h}}\Bigl |_{z_{f}} \Big[ R_{A}(M_{h}) \Big]^{k}  \Phi_{z_{c}}(\log L \mid M_{\rm{h}}) dM_{h},
\label{eqn:expandedlf}
\end{equation}

where $z_{c}$ is our calibration redshift, $R_{A}(M_{h})$ is the ratio of halo assembly times between $z_{c}$ and $z_{f}$ (dropping the variables $z_{c}$ and $z_{f}$ for brevity), and $k$ is a power index. In this form, the free parameters in our model are $\Sigma$, $\varepsilon_{DC}$ and $k$. We will simultaneously calibrate two of the free parameters ($\Sigma$ and $k$) using the QLFs at two different redshifts. For this we use the QLFs at $z = 4$ and $z=5$ given by, \citet{Akiyama_2017} and \citet{McGreer_2018}. We will see that calibration using QLFs alone is insufficient to constrain the internal duty cycle, $\varepsilon_{DC}$, thus $\varepsilon_{DC}$ is left as a free parameter. In Sec~\ref{sec:dc}, we will infer constraints for $\varepsilon_{DC}$ using additional measurements of the duty cycle from quasar clustering at high-$z$ \citep{Shen_2007, Eftekharzadeh_2015, He_2017}. For our work, we will look at 3 cases, sampling over 2 magnitudes of duty cycle: $\varepsilon_{DC} = 1, 0.1, 0.01$ corresponding to high, medium and low cases of the internal duty cycle. 

The method to derive $L_{c}(M_{h})$ under the assumption of non-zero $\Sigma$ can be challenging as this reduces down to a deconvolution problem with noise, and an exact solution is not necessarily guaranteed \citep{Ren_2020}. We approach this problem with a physically motivated systematic method: (1) Our initial guess $L'_{c}(M_{h})$ is obtained by abundance matching between the HMF with the observed QLF at $z_{c}=4$. (2) The log-normal dispersion $\Sigma$ preserves the power law slope of the faint end in the modeled QLF but broadens the bright end \citep{Cooray_2005}, thus we apply a constant scaling to $L'_{c}(M_{h})$ to renormalize the faint end of resulting QLF. (3) In \citet{Ren_2020} we show that a flattening in $L'_{c}(M_{h})$ beyond a characteristic luminosity is a sufficient approximation that provides a well behaved QLF with a bright end that matches the observed LF for a wide range of $\Sigma$ values. In this simplified approach, the process is equivalent to applying a step-function to the derivative, $\frac{dL'_{c}(M_{h})}{dM_{h}}$ and subsequently reintegrating this using any initial point that is unaffected by the step function. Here we generalise this approach by replacing the step function with a 2-parameter logistic function that we fit the bright end with. The physical motivation behind a logistic or step function is that one should expect the gradient of $L_{c}(M_{h})$ to decrease because of feedback that becomes substantial in the high halo mass regime \citep{Di_Matteo_2005, Croton_2006}, with the transition point given here as a characteristic mass, $M_{h,0}$. Therefore, we optimize the following function in order to fit our modeled QLF using Equation~\ref{eqn:expandedlf} with the observed QLF:

\begin{equation}
\dfrac{dL_{c}}{dM_{h}}(M_{h}) = A \times \dfrac{dL'_{c}}{dM_{h}}(M_{h}) \times \dfrac{1}{1 + \exp \bigl(-\gamma (\log M_{h,0} - \log M_{h}) \bigr)}.
\label{eqn:fit}
\end{equation}

$A$ is the constant scaling factor required to renormalize the QLF, $\gamma$ is a steepness parameter for the logistic function and $M_{h,0}$ is the transition point of the logistic function. For each value of $\varepsilon_{DC}$, we employ the standard $\chi^{2}$ minimization method to simultaneously calibrate all parameters, $(A, \gamma, M_{h,0})$ and $(\Sigma, k)$ over the \citet{Akiyama_2017} $z=4$ QLF and \citet{McGreer_2018} $z=5$ QLF. Additionally, we double the weight for select data points near $M^{*}$ in the \citet{Akiyama_2017} QLF (i.e. between $-26 < M_{UV} < -25.2$). This condition may seem somewhat arbitrary, but was included to penalize extremely high values of $\Sigma$ which can result in systematically underestimating the number of $M_{UV} \sim M^{*}$ quasars for our modeled $z=4$ QLF \citep{Ren_2020}. However, a calibration even without the weight adjustments will still result in good agreement with all QLF observations, albeit with a slightly sub-optimal modeled QLF at $z=4$.

\section{Results}

\subsection{$z \ge 4$ evolution of the QLF\label{sec:m4lf}}

\begin{figure}[h!]
	\centerline{\includegraphics[trim={0 1.5cm 0 0},clip, scale=0.60]{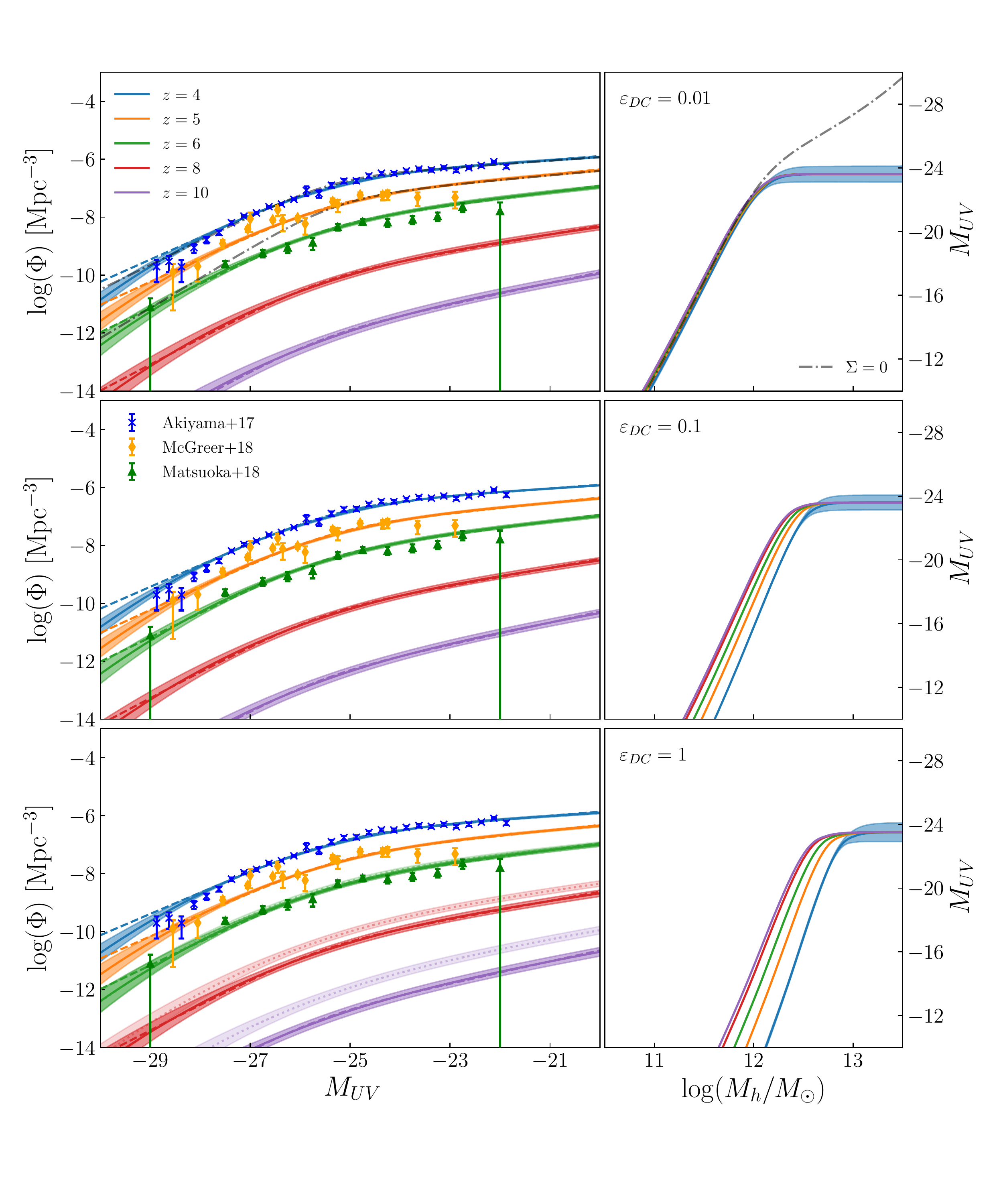}}
	\caption{\small The left panels show the modeled QLFs for $z \ge 4$ (solid colored lines) and their associated $1\sigma$ uncertainties. The right panels show the corresponding median quasar luminosity versus halo mass relation $L_{c}(M_{h})$. The $1\sigma$ uncertainty is shown for $L_{c}(M_{h})$ at $z=4$ for reference. From top to bottom show the different cases of $\varepsilon_{DC} = 0.01, 0.1$ and $1$ respectively. For the top panels, we include our model predictions if we fix $\Sigma = 0$ without calibration (dot-dashed gray lines). For the modeled QLFs for $\varepsilon_{DC} = 1$ (lower left panel), we overplot the $\varepsilon_{DC} = 0.01$ QLFs for $z\ge6$ (dotted colored lines) for comparison. The double power law fits to the curve from MCMC are also provided (dashed colored lines).}
	\label{fig:fig1}
\end{figure}

The model best fit parameters and their associated uncertainties are listed in Table~\ref{table:table0}. In Fig.~\ref{fig:fig1}, we show $L_{c}(M_{h})$ and the resulting $z\ge4$ QLF for $\varepsilon_{DC} = 1, 0.1$ and $0.01$. Here, $L_{c}(M_{h})$ at other redshifts were derived using abundance matching between the HMF and the `intrinsic' QLF, i.e. finding $L(M)$ at $z_{f}$ such that

\begin{equation}
 \int^{\infty}_{M} \dfrac{dn}{dM_{h}}\Bigl |_{z_{f}} dM_{h} = \int^{\infty}_{L} \dfrac{dn}{dM_{h}}\Bigl |_{z_{f}}\Big[ R_{A}(M_{h}) \Big]^{k}  \dfrac{dM_{h}}{dL_{c}(M_{h})}\Bigl |_{z_{c}} dL_{c}
\label{eqn:lcmh}
\end{equation}

is satisfied. The $\Sigma = 0$ case with an appropriately selected $k$ value is shown for comparison purposes in the $\varepsilon_{DC} = 0.01$ panels of Fig.~\ref{fig:fig1}. We highlight here, the impact of $\Sigma$ in reproducing the bright end of the QLF after redshift evolution. By construction, a single QLF at one redshift can be adequately fit with any $\Sigma$, although with some small deviations around $M_{UV} \sim M^{*}$ for the highest $\Sigma$ values \citep{Ren_2020}. We note that low values of $\Sigma$ underestimate the number density of the brightest quasars after redshift evolution and yields a bright end that is substantially steeper than what is observed (as seen by the dot-dashed grey line in Fig.~\ref{fig:fig1}). On the contrary, a sufficiently high $\Sigma$ will result in a flatter bright end that is consistent with the observed evolution of the QLF. Furthermore, we find that our modeled QLFs \textit{appear} to become shallower as a function of increasing $z$. This behavior of the bright end with high $\Sigma$ is better aligned with the redshift evolution of the bolometric QLF \citep{Shen_2020}, whom finds shallower slopes for higher redshifts ($z > 2.4$), compared to the evolution of the UV QLF with bright end slopes that become steeper at higher redshifts \citep{Kulkarni_2019}. However, we emphasize that the functional form of our model QLFs are not double power laws (DPL), but more akin to `broadened' Schechter functions. Any physical interpretation to the steepness of the QLF slopes can be obfuscated after forcing a double power law parameterisation to fit the QLFs, where the bright end slope, $\beta$ is sensitive to the placement of the `knee', $M^{*}$ (see Section~\ref{sec:dpl}). In this model, the evolution for the shallower bright end is primarily due to the decreased abundance of $M_{h} > M_{h,0}$ halos at higher redshifts, which leads to substantial variation in quasar luminosities within the population of the most massive halos \citep{Ren_2020}.

In Table~\ref{table:table0} we find that the best-fitting $\Sigma$ is insensitive to the choice of $\varepsilon_{DC}$, whereas $k$ is relatively sensitive to $\varepsilon_{DC}$. With the best fit parameters, the evolved QLFs are consistent with existing QLF observations at $z\ge4$ for all $\varepsilon_{DC}$, including the $z\sim6$ \citet{Matsuoka_2018} QLF that is outside of our redshift range for calibration, thus providing a validation set for our model. Since the only redshift dependent terms in Equation~\ref{eqn:expandedlf} are the HMF and the halo assembly times, our model therefore suggests that the evolution of the overall quasar density is driven by the HMF, with some possible contribution from $R_{A}(M_{h})$ and its associated power index $k$, depending on $\varepsilon_{DC}$. One interpretation for $R_{A}(M_{h})$ is to associate this quantity as a proxy for the contribution of merger-driven growth to the SMBH, which can be expected to account for some part in the SMBH growth history \citep{Marshall_2020}. We see in Equation~\ref{eqn:lcmh} that the effect of $R_{A}(M_{h})$ is to provide a boost in $L_{c}(M_{h})$. From Fig.~\ref{fig:fig1}, we find that this boost can be significant, up to $2$ mag, for high values of $\varepsilon_{DC}$ (e.g. $\varepsilon_{DC} = 1$), and marginal, with corresponding $z\ge4$ $L_{c}(M_{h})$ curves within $1\sigma$ of each other, for smaller values of $\varepsilon_{DC}$ (e.g. $\varepsilon_{DC} = 0.01$).

While our model contains a number of parameters, the calibration step effectively fixes all of them except for $\varepsilon_{DC}$. Furthermore, from the left panels of Fig.~\ref{fig:fig1}, we see that the model results are only weakly depending on the choice of $\varepsilon_{DC}$. It is noteworthy that a constant $\varepsilon_{DC}$ used in our model is sufficient to fully reproduce all observed QLFs at $z \ge 4$. For all $\varepsilon_{DC}$ considered, the modeled QLFs at $z \le 7.5$ are within $2\sigma$ from each other, with the most difference at high $z$. This potentially offers the novel possibility of using a future determination of the QLF at $z\gtrsim 7.5$ as an independent constraint for the duty cycle, distinct from derivations of the duty cycle based on clustering measurements. We explore further the effect of $\Sigma$ and $\varepsilon_{DC}$ for measurements of the duty cycle in Section~\ref{sec:dc}.

Fig.~\ref{fig:fig1} also allows us to investigate how the characteristic mass scale $M_{h,0}$ in $L_{c}(M_{h})$ evolves as a function of $z$. We find that the evolution of $M_{h,0}$ depends on $\varepsilon_{DC}$, with $\varepsilon_{DC} = 0.01$ showing no visible redshift evolution and $M_{h,0} \sim 10^{12.1} M_{\odot}$, while the $\varepsilon_{DC} = 1$ case displays a range of $M_{h,0} \sim 10^{12.6} - 10^{13} M_{\odot}$ for $4 \le z \le 10$. A more consistent feature is that of a characteristic luminosity $L_{c,0}$ that is constant for all $\varepsilon_{DC}$ and $z$. If we make the assumption that the brightest quasars are radiating at Eddington luminosity, then our model implies a characteristic SMBH mass. One can then interpret $L_{c,0}$ as an indication that the main mechanism for the self-regulation of SMBH growth in the average quasar is due to the quasar itself, as opposed to the radio mode accretion that is typically expected to quench star formation in massive halos with $M_{h} > 10^{12} M_{\odot}$ \citep{Croton_2006}. We stress that these feedback modes are not mutually exclusive with each other and that an individual quasar can be affected by either mechanism. Furthermore, we note that even for our maximal $\varepsilon_{DC} = 1$ case, the shift in $M_{h,0}$ is relatively small, just $\Delta M_{h,0} = 0.4$, thus it does not exclude the possibility that the main mode of self-regulation for the average quasar is still due to radio mode feedback. For smaller values of $\varepsilon_{DC}$, the model begins to completely lose the ability to discriminate between the primary mechanism that self-regulates SMBH growth, whether it is a characteristic mass $M_{h,0}$, luminosity $L_{c,0}$, or a combination thereof.

\subsection{$z \le 4$ evolution of the QLF\label{sec:l4lf}}

\begin{figure}[h!]
	\centerline{\includegraphics[trim={0 1.5cm 0 0},clip, scale=0.60]{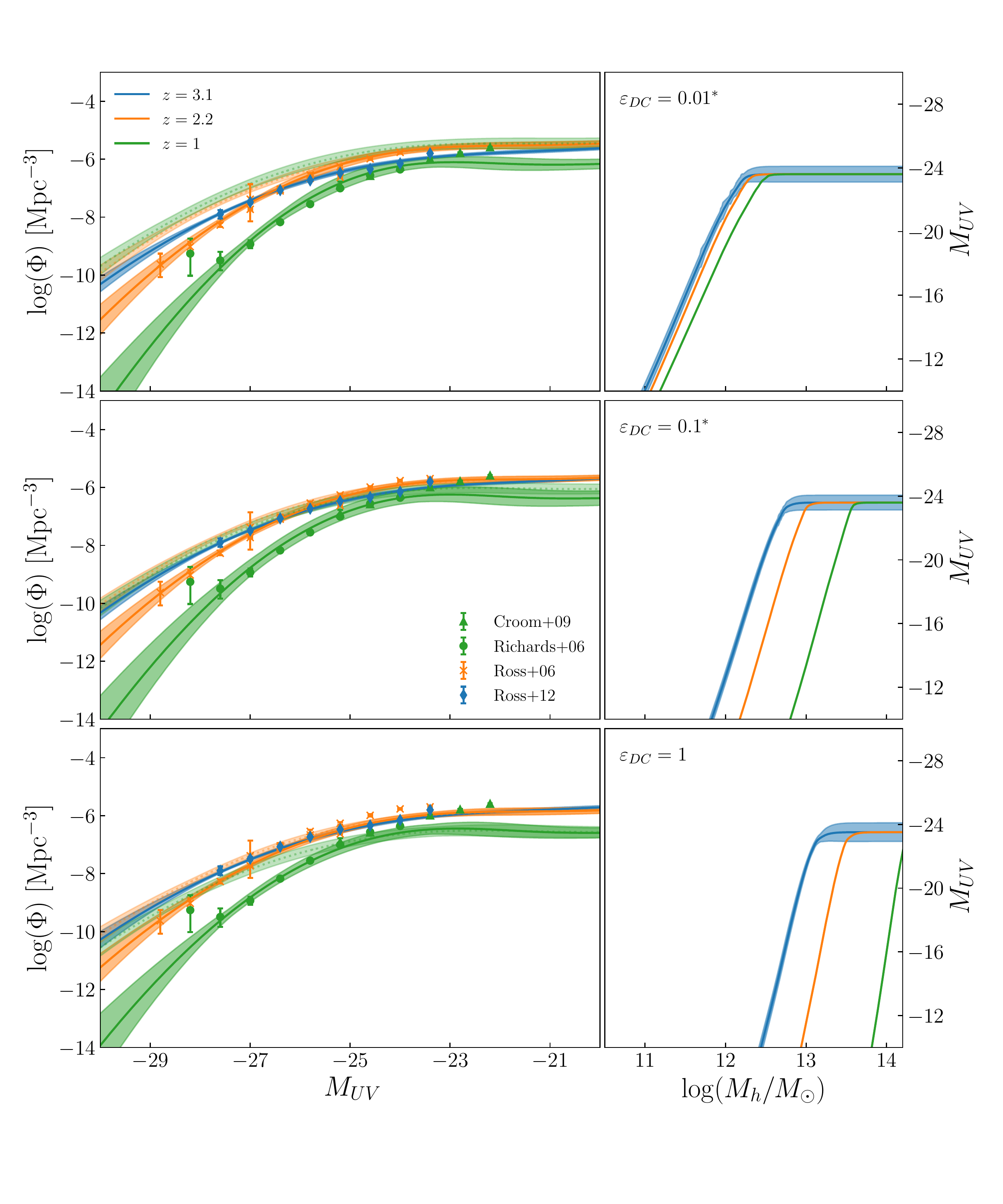}}
	\caption{\small As in Fig.~\ref{fig:fig1}, but for $z \le 4$. The $z = 2.2(1.0)$ QLFs for all $\varepsilon_{DC}$ requires a lower dispersion, $\Delta\Sigma \sim -0.09(-0.19)$ in order to compensate for the overproduction of bright quasars. The $z = 1$ QLF in the $\varepsilon_{DC} = 0.01(0.1)$ scenario also required a scaling in $\varepsilon_{DC}$ by a factor of $0.2(0.5)$ in order to reproduce the overall drop in quasar number density. The original curves as produced by the fiducial model are shown in the dotted colored lines.}
	\label{fig:fig2}
\end{figure}

We investigate the applicability of this model for $z \le 4$ as well (i.e. below the redshift range at which the parameters were calibrated). In Fig.~\ref{fig:fig2}, we show our fiducial model predictions for $z = 3.1$ against the observed QLF resulting in an excellent agreement to observations for all $\varepsilon_{DC}$. However, as redshift decreases, a small evolution in $\Sigma$ is necessary to prevent the over-representation of bright quasars. For example, the $z = 2.2$ QLF requires a tighter scatter with $\Delta \Sigma \sim -0.09$ relative to the nominal constant $\Sigma$ used for higher $z$ for all values of $\varepsilon_{DC}$. As redshift decreases, $\Delta \Sigma$ grows, e.g. we find $\Delta \Sigma \sim -0.19$ at $z=1.0$. Physically, it is not implausible to expect a tighter scatter in $L_{c}(M_{h})$ at lower redshift. For example, mass averaging over numerous successive merging events could restrict the spread in SMBH masses \citep{Peng_2007}, therefore resulting in a reduced dispersion in $\Sigma$. We overlay the fiducial $\Sigma$ predictions in Fig.~\ref{fig:fig2} as dotted colored lines to highlight the extent of the overproduction of bright quasars if $\Sigma$ was not corrected for in the low-$z$ regime. 

The low redshift evolution of the QLF in our model is also impacted by the choice of the power law index parameter $k$. Since $R_{A}(M_{h}) < 1$ for $z_{f} < z_{c}$, a high $k$ damps the effective growth of the QLF compared to what would be expected simply from the evolution of the HMF. For substantial values of $k$, this may even lead to a decrease of the overall quasar number density going to lower redshifts, as seen in the $z=2$ to $z=1$ QLFs for $\varepsilon_{DC} = 1$ in Fig.~\ref{fig:fig2}. However, we stress that $\varepsilon_{DC} = 1$ is an extreme scenario where the large $k$ can lead to possibly unphysical consequences, such as a significant evolution in the SMBH - host galaxy mass relation which is not expected theoretically nor inferred from observations. In fact, the SMBH - host galaxy mass relation is broadly expected to follow the evolution in $L_{c}(M_{h})$ since neither the stellar efficiency parameter \citep{Behroozi_2012} or the Eddington ratio distribution \citep{Kollmeier_2006} show strong evidence of redshift evolution up to $z \sim 4$. Both observations \citep{Decarli_2010} and simulations \citep{Volonteri_2016, Marshall_2020} support only a mild evolution in the SMBH - host galaxy mass relation. However, our modeling finds that any evolution can be largely suppressed by assuming a lower intrinsic duty cycle $\varepsilon_{DC} \le 0.1$, which yields a lower $k$. On the other hand, assuming $k << 1$ also leads to difficulties in reproducing the observed QLF at $z = 1$ under fiducial model assumptions, due to an over-representation of fainter quasars. In this scenario, we find that our model is able to approximately match the $z = 1$ QLF only if we allow the duty cycle $\varepsilon_{DC}$ at $z=1$ to scale by a factor of $0.2(0.5)$ for starting values $\varepsilon_{DC} = 0.01(0.1)$. A decrease in the value of $\varepsilon_{DC}$ at $z = 1$ is plausible through observations of shorter quasar lifetimes, $t_{Q} \sim 10^{7}$ Myr at $z\lesssim 1$ \citep{Porciani_2004, Padmanabhan_2009}, corresponding to a duty cycle of $\sim t_{Q}/t_{H} \sim 10^{-3}$.

It is interesting that the assumption of a redshift-independent duty cycle yields a very good match to observations throughout a large redshift range, well beyond the calibration interval. In fact, as in the $z\ge 4$ case, we find that the evolution in the overall number density for $z\le 4$ can be determined from the combined evolution of the HMF and $R_{A}(M_{h})$. Thus our model suggests that the ensemble phenomenology of dark matter halos in combination with AGN feedback alone is sufficient to fully describe the evolution of the QLF. There is no requirement for any change in the global ISM conditions (such as a global depletion of gas or a redshift-dependent cosmic dust extinction), at least for $z\ge2$.

One consistent feature that our model is unable to fully replicate, even after allowing an evolution for $\Sigma$ and $\varepsilon_{DC}$, is the faint end of the $z \le 2$ QLF. In fact, our model routinely underestimates the distribution of fainter quasars at $z = 1$, reaching up to $\sim 1$ order of magnitude difference in counts for faint $M_{UV} = -22$ objects. We attribute this discrepancy to a limitation of our modeling method, specifically to our choice to calibrate the QLF at $z = 4$ and $z = 5$. In reality, at $z \le 2$, we may expect a population of AGN with massive SMBHs ($M_{\rm BH}> 10^{7} M_{\odot}$) that are quiescently accreting in radio-mode which ultimately contributes to the faint end of the QLF \citep{Hirschmann_2014}. The calibration at high $z$ is not sensitive to this secondary population as the number of these objects is expected to be very low at high $z$. Interestingly, the inability to account for this feature is beneficial for our modeling of the QLF across most $z$, but particularly at high $z$, as this allows us to take advantage of using a simple flattening of $L_{c}(M_{h})$ at the high mass end to capture the general qualities in the evolution of the QLF. However, the natural drawback is that our model does not account for the population of AGN inside massive halos that may play a larger role at lower redshifts.

\subsection{Measured (observed) Duty Cycle\label{sec:dc}}

The effect of quasar luminosity scatter $\Sigma$ can impact measurements of the duty cycle derived from quasar-quasar or quasar-galaxy clustering, as $\Sigma$ decreases median halo mass hosting highly luminous quasars \citep{Padmanabhan_2009, Shankar_2010, Ren_2020}. Our modeling framework presents the opportunity to quantitatively investigate how our calibrated $\Sigma$ can impact the duty cycle that would be inferred in a mock survey as a function of $\varepsilon_{DC}$ and a survey's limiting magnitude. We employ the standard definition of the duty cycle (e.g. as in \citealt{Eftekharzadeh_2015, He_2017}) defined as,

\begin{equation}
\varepsilon_{\rm{measured}} = \dfrac{n_{\rm QSO}}{V \int^{\infty}_{\bar M_{\rm{min}}}\frac{dn}{dM_{h}} dM_{h}},
\label{eqn:dc}
\end{equation}

where $n_{\rm QSO}$ is the number of visible quasars in a given survey volume, $V$. $\bar{M}_{\rm min}$ is the median halo mass from the set of observable (i.e. sufficiently bright) quasars and $\frac{dn}{dM_{h}}$ is the HMF. The survey volume $V$ is only important in determining the variation of $\varepsilon_{\rm{measured}}$. Thus, instead of a fixed survey area, we use a fixed volume $100$Gpc$^{3}$ to simply represent an arbitrary wide-area survey. For a mock `survey', we sample the HMF, calculate quasar luminosities through the stochastic relationship $L_{c}(M_{h})$, and solve Equation~\ref{eqn:dc} assuming some limiting magnitude, $M_{\rm lim}$. We investigate two distinct survey arrangements over $2 \le z \le 10$: (1) First, we use a limiting magnitude $M_{\rm lim} = -22.9$, roughly consistent with the lower luminosity limit from both of the quasar samples used in the $z < 3$ SDSS-III BOSS catalog from \citet{Eftekharzadeh_2015} and the low-luminosity sample from the $z\sim 4$ HSC wide-layer catalog of \citet{He_2017}; For (2) we assume a brighter magnitude limit of $M_{\rm lim} = -27$ to be more representative of the quasar samples used in the SDSS DR5 catalogue from \citet{Shen_2007} and also the high-luminosity sample of \citet{He_2017}.

\begin{figure}[h!]
	\centerline{\includegraphics[angle=-00, scale=0.60]{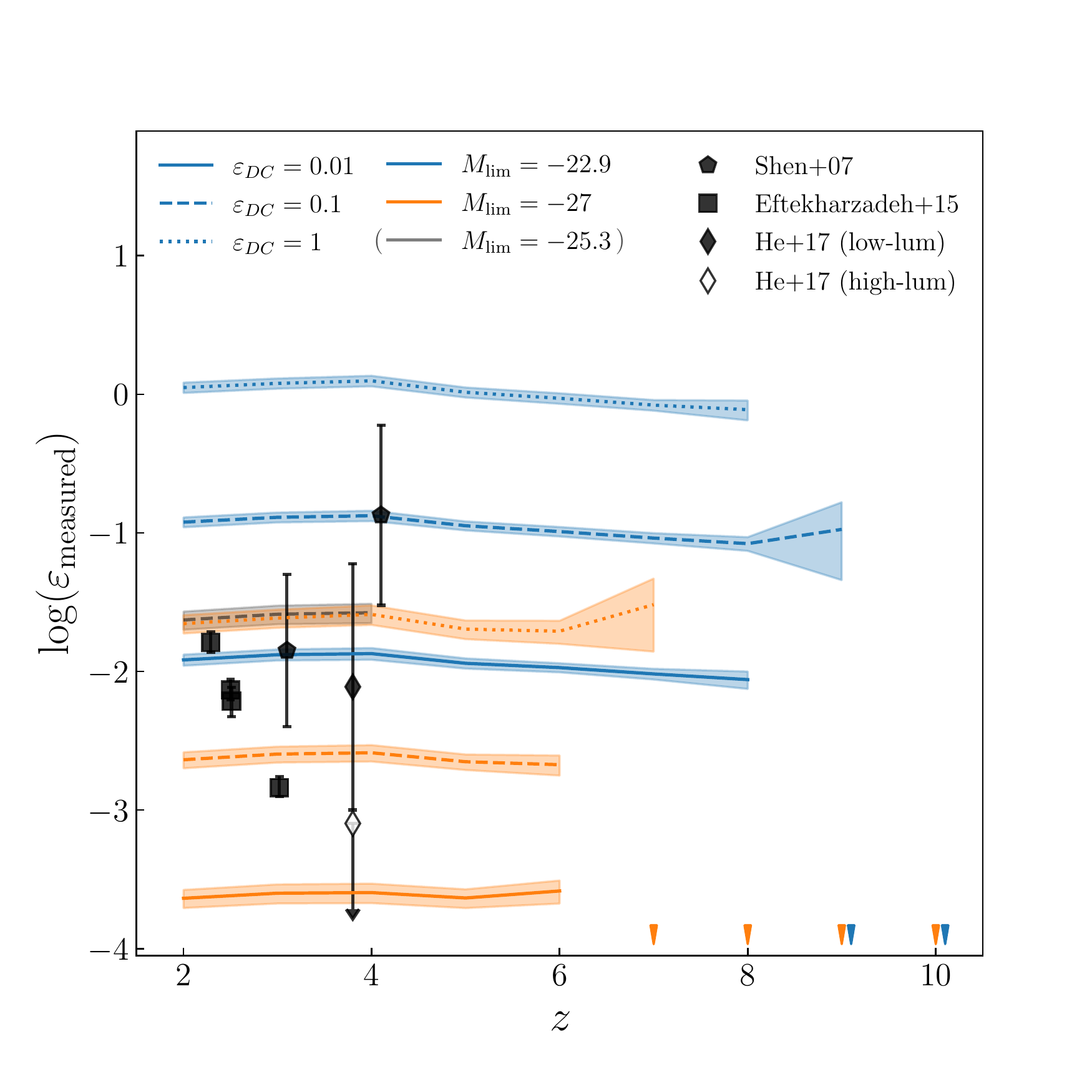}}
	\caption{\small Measured duty cycle as a function of $z$ and $\varepsilon_{DC}$. Black points are observations. The blue lines represent mock `surveys' sampled at various redshifts with $V = 100$Gpc$^{3}$ and $M_{\rm lim} = -22.9$. The orange lines represent mock surveys with the same survey volume as the blue lines, but $M_{\rm lim} = -27$. The grey line represent mock surveys with the same survey volume, but for $z\le4$ and $M_{\rm lim} = -25.3$ and was shown to indicate the lower limit of $\varepsilon_{DC}$ with the observations of \citet{Shen_2007}. Values of $\varepsilon_{DC} = 0.01, 0.1, 1$ are separated by solid, dashed and dotted line styles respectively. Each mock survey was repeated $10$ times for each redshift range considered. The wedges at the bottom edge of the plot are an indication that out of the $10$ mock surveys, there were insufficient recorded quasars to collect any meaningful statistics for the survey parameters and redshifts considered (blue or orange symbols).}
	\label{fig:fig3}
\end{figure}

We conduct $10$ mock surveys for each of the scenarios presented above. In Fig.~\ref{fig:fig3}, we show the measured duty cycle, $\varepsilon_{\rm{measured}}$ as a function of redshift and $\varepsilon_{DC}$. In all cases, we find that $\varepsilon_{\rm{measured}}$ is relatively independent of $z$. Furthermore, we note that there is an overall good agreement between $\varepsilon_{DC}$ and $\varepsilon_{\rm{measured}}$ on the provision that the limiting magnitude is sufficiently sensitive (blue lines of Fig.~\ref{fig:fig3}). This surprising result is contrary to the earlier expectation that the quasar luminosity scatter, $\Sigma$ should reduce the measured clustering and therefore should result in a lower $\varepsilon_{\rm{measured}}$. While we do expect the general qualitative impact of changing $\Sigma$ on clustering to hold, we predict that given the range of $\Sigma$ and the above specific survey parameters, the significantly larger population of fainter quasars with a smaller spread of possible halo masses is sufficient to overcome the dilution of clustering strength from $\Sigma$ to recover the internal duty cycle as the measured duty cycle. In contrast, we see that a shallower magnitude limit $M_{\rm lim} = -27$ results in a reduced $\varepsilon_{\rm{measured}}$ from the impact of $\Sigma$ (orange lines of Fig.~\ref{fig:fig3}). In this scenario, $M_{\rm lim}$ is much brighter than the magnitude corresponding to $L_{c}(M_{h,0})$, which is determined by $\Sigma$, thus any detected quasar is biased towards having a halo mass close to $M_{h,0}$. This effectively suppresses the measured duty cycle determined by Equation~\ref{eqn:dc}. Thus, in order to obtain a robust measurement of the duty cycle, our modeling implies that the limiting depth of a survey needs to be at least comparable to $L_{c}(M_{h,0})$.

Under the survey conditions of \citet{Eftekharzadeh_2015} and \citet{He_2017}, we find that that our modeling is consistent with their results assuming an internal duty cycle of a few percent, i.e. $\varepsilon_{DC} \sim 0.01$. Furthermore, this modeling prescription offers an explanation for the drastically lower duty cycle ($\varepsilon_{\rm{measured}} \sim 8 \times 10^{-4}$) determined from the high luminosity $z\sim 4$ quasar sample of \citet{He_2017}. As the high luminosity quasar sample used in \citet{He_2017} spans a magnitude range between $-24 \lesssim M_{UV} \lesssim -28$, we speculate that observed quasars are overwhelmingly biased in residing within lower mass halos. In contrast, \citet{Shen_2007} determine a measured duty cycle of  $\varepsilon_{\rm{measured}} \sim 0.03 - 0.6$ from a $z > 3.5$ quasar sample which includes objects $M_{UV} < -25.3$ using the band conversion in \citet{Ross_2013}. The upper end of this value is largely discrepant to the other measurements by \citet{Eftekharzadeh_2015} and \citet{He_2017} with the difference in measurements being attributed to the treatment of the large scale ($ >30$ Mpc) data points. If we were to take the measurements of \citet{Shen_2007} at face value for constraining the value $\varepsilon_{DC}$, then this would imply an internal duty cycle that is larger than the measured duty cycle, as the survey's limiting depth is brighter than $L_{c}(M_{h,0})$, thus exacerbating the existing discrepancy in duty cycle values. Our modeling finds an equivalent lower limit of $\varepsilon_{DC} \sim 0.1$ using the survey parameters of \citet{Shen_2007}.

\subsection{Forecasts for the Nancy Grace Roman Telescope\label{sec:rst}}

\begin{figure}[h!]
	\centerline{\includegraphics[angle=-00, scale=0.60]{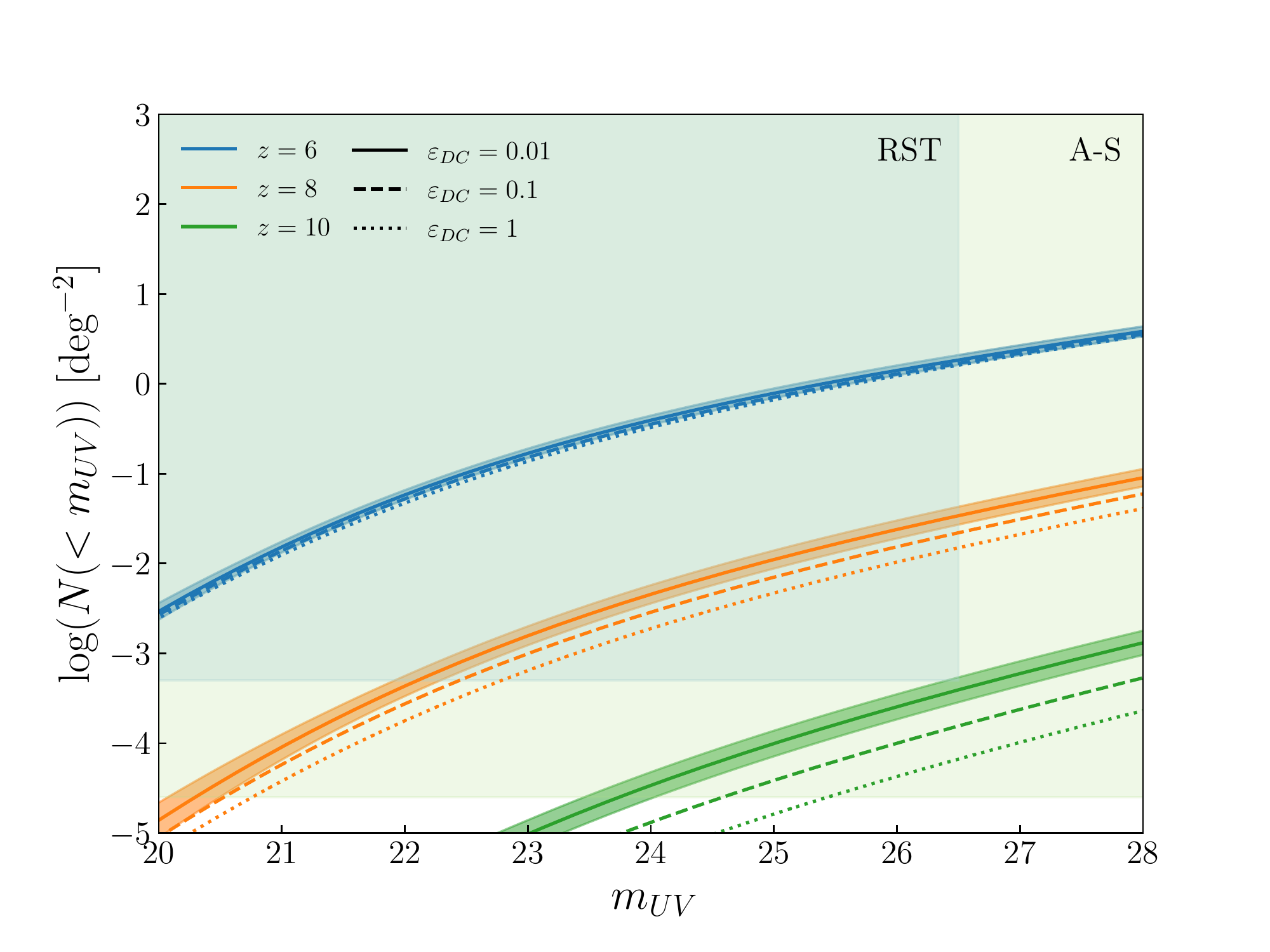}}
	\caption{\small Predicted cumulative number counts of quasars with $< m_{UV}$ per square degree at different redshifts with $\varepsilon_{DC} = 0.01$ (solid lines), $\varepsilon_{DC} = 0.1$ (dashed lines) and $\varepsilon_{DC} = 1$ (dotted lines). The indicative $1\sigma$ uncertainties is shown for the $\varepsilon_{DC} = 0.01$ case (solid colored lines). The darker shaded region corresponds to the wide-area mission for the RST ($\sim 2000$ deg$^{2}$) and the lighter region is the all-sky coverage ($\sim 40000$ deg$^{2}$). The intersection between the curve and the bottom edge of the shaded regions indicate the brightest quasar one can expect to find under the corresponding survey size. Similarly, the difference between the bottom edge of the shaded regions to the curve provides the logarithm of the number of objects at the respective $m_{UV}$ value.} 
	\label{fig:fig4}
\end{figure}

In Fig.~\ref{fig:fig4} we use our model to provide a forecast accessible by the proposed wide-field Nancy Grace Roman Telescope (RST) mission ($\sim 2000$ deg$^{2}$) as well as the total number of quasars expected over the all-sky area ($\sim 40000$ deg$^{2}$). We find that RST is easily capable of finding $z\sim8$ quasars assuming an apparent magnitude limit of $m_{UV} = 26.5$, with an expected total of $N \gtrsim O(10^{1})$ detected objects for all values of $\varepsilon_{DC}$. Additionally, at the same magnitude depth, we predict that the earliest quasar observable in the all-sky area is $z\sim 10(11)$ for $\varepsilon_{DC} = 1(0.01)$.

\subsection{Double Power Law fits\label{sec:dpl}}

As standard in the literature, we provide the double power law fits (DPL) for the modeled $z \ge 3$ QLF in Table~\ref{table:table1}. The DPL fits are represented as the dashed lines in Fig.~\ref{fig:fig1}. The DPL function is given by

\begin{equation}
\phi(M_{UV}) = \dfrac{\phi^{*}}{10^{0.4(\alpha + 1)(M_{UV} - M^{*})} + 10^{0.4(\beta + 1)(M_{UV} - M^{*})}},
\label{eqn:dpl}
\end{equation}

where $\phi^{*}$ is the normalization, $\alpha$ is the faint end slope, $\beta$ is the bright end slope and $M^{*}$ is the characteristic break magnitude. To calculate the best fit parameters we compute the posterior probability distribution using the Monte Carlo Markov Chain technique (MCMC; \citealt{Foreman_Mackey_2013}) with the standard likelihood, $\mathcal{L} \sim \exp(-\chi^{2})$ over a luminosity range of $M_{UV} \in [-30, -20]$. We assume uniform priors for all parameters and enforce a DPL shape with the constraint such that $\alpha$ and $\beta$ do not overlap prior domains.

We note that the DPL fit parameters $\alpha, \beta$ and $M^{*}$ are fully consistent with each other in $\varepsilon_{DC}$, while the normalization $\phi^{*}$ is the sole parameter to vary with $\varepsilon_{DC}$. The best fit parameters suggest a steepening of the faint end slope $\alpha$ in redshift, following the evolution of the faint end slope of the HMF. In contrast, we find a very mild steepening of the bright end slope $\beta$ in redshift. However, we strongly caution the significance of this finding as the evolutionary trend in $\beta$ also correlates with an evolution of the DPL break point $M^{*}$. Since our modeled QLF is effectively a Schechter shape (from the HMF) convolved with a log-normal kernel, we can expect a fit that sets $M^{*}$ beyond the `knee' of the modeled QLF will result in a steeper $\beta$ `fit'. Observational determinations of $(M^{*}, \beta)$ seem to be consistent with this correlation \citep{Kulkarni_2019}. Therefore, we find that it is difficult to associate the steepening trend of the DPL bright end slope to any physical mechanism, especially in our case where the trend is mild. Furthermore, an inspection of the QLF in Fig.~\ref{fig:fig1} suggests the contrary, i.e. the bright end of the HMF appears to flatten due to the broadening from quasar luminosity scatter $\Sigma$. 

With respect to the calibration QLFs, the DPL parameters derived for the $z=4$ QLF show good agreement with \citet{Akiyama_2017}, except for a slightly underestimated $\beta$. However, the calibrated QLF is still fully consistent with the entire set of observed data points. In contrast, our DPL parameters does not agree with the fit for the $z=5$ QLF by \citet{McGreer_2018}. The discrepancy in the latter is attributed to the choice of parameterization in fixing $\beta = -4.0$. We find that the DPL fits systematically overproduces UV-bright quasars compared to the results of our full modeling. This is expected as from the Schechter nature of the modeled QLF, because the bright end retains its exponential shape even after accounting for quasar luminosity scatter $\Sigma$. However, we note that the differences between model and fit are only restricted to the brightest objects and thus are still well within current observational uncertainties. 

\section{Conclusions} \label{sec:conclus}

It can be expected that the number density evolution with redshift for objects that reside inside dark matter halos, such as quasars, will follow the general trends of the HMF. For example, semi-empirical models for galaxies have shown that accurate determinations of the galaxy LF can be constructed based on the evolution of the HMF \citep{Trenti_2010, Mason_2015}. Following this intuition, we construct a simple semi-empirical framework to model the evolution of the UV QLF across $z$. Our modeling method takes advantage of the recent determinations of the QLF at $z = 4-5$ to calibrate free parameters, resulting in a model that solely depends on the internal duty cycle, $\varepsilon_{DC}$. We summarize the key features of our model below:

\begin{itemize}

\item The bright and the faint ends of the QLF are linked to the Schechter shape of the HMF itself. Here, the bright end shape results from a convolution of the exponential profile of a Schechter with some log-normal scatter, whereas the QLF faint end is simply a power-law like the HMF, since a power law functional form is insensitive to scatter \citep{Cooray_2005}. Our model finds that a constant quasar luminosity scatter, $\Sigma \sim 0.6$, irrespective of $\varepsilon_{DC}$, is able to reproduce the bright end of the QLF at $z \ge 3$.

\item There is a direct relation of the evolution of the QLF with the evolution of the HMF. We find that the redshift evolution of the QLF at $z \ge 3$ can be derived from the evolution of the HMF (and of its associated halo assembly time which also enters in our model) together with the assumption of a constant $\varepsilon_{DC}$. The choice of $\varepsilon_{DC}$ effectively sets the contribution from merger-related activity, as described in our model with the inclusion of a factor proportional to the halo assembly time to some power, $k$. The significance of this latter factor grows with $\varepsilon_{DC}$, with $\varepsilon_{DC} = 0.01$ requiring a small-to-marginal contribution from merger-related activity, $k = 0.25$, up to $k = 2.65$ for $\varepsilon_{DC} = 1$. Discrimination between different values of $\varepsilon_{DC}$ may be possible thanks to future observations of the QLF at $z\gtrsim7.5$.

\item Our model is unable to predict the $z \le 2$ QLFs without changes in $\Sigma$ and $\varepsilon_{DC}$. Specifically, we require $\Delta \Sigma \sim 0.09(0.19)$ at $z \sim 2(1)$ to compensate for the overestimation of bright quasars and a reduction in $\varepsilon_{DC} = 0.05(0.002)$ at $z = 1$ for initial values $\varepsilon_{DC} = 0.1(0.01)$ to reproduce the drop in overall quasar number density. By adopting these additional changes, we are able to recover the QLF across this extended redshift range, suggesting that additional degrees of freedom are needed as the physical conditions for BH growth evolve across cosmic time.

\end{itemize}

Overall, we introduce a semi-empirical model that meaningfully highlights key physical properties of SMBH growth, and demonstrates the capacity to predict the evolution of quasar properties over redshift under minimal assumptions. We find that comprehensive measurements of clustering over redshift along with accurate determinations of the QLF for $z \ge 6$ are necessary ingredients to discriminate between values of $\varepsilon_{DC}$, the free parameter in our model. Despite the uncertainty in $\varepsilon_{DC}$, the outlook for next generation wide-area surveys at high $z$ seems promising. Even assuming the worst case scenario with maximum $\varepsilon_{DC}$, our model predicts $N \gtrsim 10$, bright $m_{UV} < 26.5$ quasars at $z\sim 8$ for the $\sim 2000$deg$^{2}$ wide-area RST mission. With the same parameters, our model also predicts the earliest observable quasar for that telescope to be at $z\sim10$ once the all sky survey is completed.

\acknowledgements
We thank the anonymous referee for their insightful comments and recommendations. This research was conducted by the Australian Research Council Centre of Excellence for All Sky Astrophysics in 3 Dimensions (ASTRO 3D), through project number CE170100013. K.R is additionally supported through the Research Training Program Scholarship from the Australian Government and the Postgraduate Writing-up Award granted by the David Bay Fund.

The binned and band-corrected QLF data points for $z \le 4$ was computed using the \citet{Kulkarni_2019} homogenized AGN catalogue. The respective github repository is available at: \texttt{https://github.com/gkulkarni/QLF}.

Data products from this manuscript can be found in the github repository: \\ \texttt{https://github.com/renkeven/QuasarEvolutionData}.

\begin{deluxetable}{crrrrr}
\tablewidth{0pt}
\tablecaption{Model best-fit parameters}
\label{table:table0}
\tablehead{\colhead{$\varepsilon_{DC}$}&\colhead{$\Sigma$}&\colhead{$k$}&\colhead{$M_{h,0}$}&\colhead{$\gamma$}&\colhead{$A$}\\}
\startdata
$0.01$	&$0.61 \pm 0.03$&	$0.25 \pm 0.09$		&	$12.21 \pm 0.05$		&	$10.53\pm 0.40$		&$0.12 \pm 0.02$ \\ 
$0.1	$	&$0.61 \pm 0.03$&	$1.45 \pm 0.08$		&	$12.62 \pm 0.03$		&	$11.30 \pm 0.60$	&$0.12 \pm 0.01$	\\
$1.0	$	&$0.63 \pm 0.03$&	$2.65 \pm 0.09$		&	$12.94 \pm 0.03$		&	$10.39 \pm 0.46$	&$0.02 \pm 0.02$	\\
\enddata
\end{deluxetable}

\begin{deluxetable}{crrrrr}
\tablewidth{0pt}
\tablecaption{Determinations of Double Power Law Parameters for the QLF with their $1\sigma$ uncertainties}
\label{table:table1}
\tablehead{\colhead{Redshift}&\colhead{$\varepsilon_{DC}$}&\colhead{$\alpha$}&\colhead{$\beta$}&\colhead{$M^{*}$}&\colhead{$\log(\phi^{*})$}\\}
\startdata
$3.1	$	&0.01&	$-1.28 \pm 0.02$		&	$-3.03 \pm 0.04$		&	$-25.51^{+0.08}_{-0.07}$			&$-6.20 \pm 0.03$	\\ 
$	$	&0.1&	$-1.27 \pm 0.01$		&	$-3.01^{+0.03}_{-0.04}$		&	$-25.48 \pm 0.07$			&$-6.24 \pm 0.03$	\\
$	$	&1.0&	$-1.29 \pm 0.02$		&	$-2.93 \pm 0.04$		&	$-25.40 \pm 0.08$			&$-6.27 \pm 0.03$	\\ \hline
$4	$	&&	$-1.34 \pm 0.01$		&	$-2.93 \pm 0.02$		&	$-25.31 \pm 0.04$			&$-6.62 \pm 0.02$	\\ 
$	$	&`` ''&	$-1.32 \pm 0.01$		&	$-2.89 \pm 0.02$		&	$-25.23 \pm 0.04$			&$-6.58 \pm 0.01$	\\
$	$	&	&$-1.36 \pm 0.01$		&	$-2.85 \pm 0.03$		&	$-25.24 \pm 0.05$			&	$-6.60 \pm 0.02$	\\ \hline
$5	$	&&	$-1.45 \pm 0.01$		&	$-2.99 \pm 0.03$		&	$-25.33 \pm 0.06$			&$-7.31 \pm 0.03$	\\ 
$	$	&`` ''&	$-1.45 \pm 0.01$		&	$-2.99 \pm 0.03$		&	$-25.34 \pm 0.06$			&$-7.31 \pm 0.03$	\\
$	$	&	&$-1.46 \pm 0.01	$		&	$-2.92 \pm 0.04$		&	$-25.25 ^{+0.08}_{-0.07}$			&	$-7.29 \pm 0.03$	\\ \hline
$6	$	&&	$-1.56 \pm 0.02$		&	$-3.11 \pm 0.05$		&	$-25.44^{+0.11}_{-0.10}$			&$-8.13 \pm 0.05$	\\ 
$	$	&`` ''&	$-1.57 \pm 0.02$		&	$-3.11 \pm 0.05$		&	$-25.47^{+0.11}_{-0.10}$			&$-8.19 \pm 0.05$	\\
$	$	&	&$-1.58 \pm 0.02	$		&	$-3.06 \pm 0.06$		&	$-25.42 \pm 0.13$			&	$-8.22 \pm 0.06$	\\ \hline
$7	$	&&	$-1.66 \pm 0.03$	&	$-3.20 \pm 0.07$		&	$-25.51 \pm 0.15$			&$-9.02 \pm 0.08$	\\
$	$	&`` ''&	$-1.68 \pm 0.03$		&	$-3.20 \pm 0.07$		&	$-25.56 \pm 0.15$			&$-9.16 \pm 0.08$	\\
$	$	&	&$-1.69 \pm 0.03$		&	$-3.15 \pm 0.08$		&	$-25.52 \pm 0.18$			&	$-9.25 \pm 0.09$	\\  \hline
$8	$	&&	$-1.75 \pm 0.03$		&	$-3.27 \pm 0.09$		&	$-25.56^{0.21}_{0.20}$			&$-9.97 \pm 0.11$	\\
$	$	&`` ''&	$-1.77 \pm 0.03$		&	$-3.27^{+0.08}_{-0.09}$		&	$-25.62^{+0.20}_{-19}$			&$-10.20 \pm 0.11$	\\
$	$	&	&$-1.78 \pm 0.04$		&	$-3.22 ^{+0.09}_{-0.10}$	&	$-25.59 ^{+0.24}_{-0.23}$		&	$-10.37 \pm 0.13$	\\ \hline
$9	$	&&	$-1.84 \pm 0.04$		&	$-3.33^{+0.10}_{-0.11}$		&	$-25.60^{+0.26}_{-0.25}$			&$-10.95 \pm 0.15$	\\ 
$	$	&`` ''&	$-1.86 \pm 0.04$		&	$-3.33^{+0.10}_{-0.11}$		&	$-25.67^{+0.25}_{-0.24}$			&$-11.30 \pm 0.14$	\\
$	$	&	&$-1.87 \pm 0.07	$		&	$-3.28 ^{+0.11}_{-0.12}$	&	$-25.63 ^{+0.29}_{-0.28}$		&	$-11.56 \pm 0.17$	\\ \hline
$10	$	&&	$-1.92 \pm 0.05$		&	$-3.39^{+0.12}_{-0.13}$		&	$-25.63^{+0.32}_{-0.30}$			&$-11.98 ^{+0.19}_{-0.18}$	\\ 
$	$	&`` ''&	$-1.94^{+0.05}_{-0.04}$		&	$-3.39^{+0.11}_{-0.12}$		&	$-25.71^{+0.30}_{-0.28}$			&$-12.44^{+0.18}_{-0.17}$\\
$	$	&	&$-1.95 \pm 0.05$		&	$-3.34 ^{+0.12}_{-0.13}$	&	$-25.67 ^{+0.35}_{-0.33}$		&	$-12.80 \pm 0.21$	\\
\enddata
\end{deluxetable}

\bibliographystyle{aasjournal}
\bibliography{qsoevo}
\end{document}